\documentclass{aip-cp}

\usepackage[numbers]{natbib}
\usepackage{rotating}
\usepackage{graphicx}
\newcommand{\ba}{\begin{eqnarray}}
\newcommand{\ea}{\end{eqnarray}}
\begin{document}

\title{Symmetries and Order in Cluster Nuclei}

\author[aff1]{Roelof Bijker}
\eaddress{bijker@nucleares.unam.mx}

\affil[aff1]{Instituto de Ciencias Nucleares, 
Universidad Nacional Aut\'onoma de M\'exico,\\ 
A.P. 70-543, 04510 Cd. de M\'exico, M\'exico}

\maketitle

\begin{abstract}
It is shown that the rotational band structure of the cluster states in $^{12}$C 
and $^{16}$O can be understood in terms of the underlying discrete symmetry that 
characterizes the geometrical configuration of the $\alpha$-particles, {\it i.e.} 
an equilateral triangle for $^{12}$C, and a regular tetrahedron for $^{16}$O. 
The structure of rotational bands provides a fingerprint of the underlying 
geometrical configuration of $\alpha$-particles. Finally, some first results are 
presented for odd-cluster nuclei. 
\end{abstract}

\section{INTRODUCTION}

The concept of symmetries has played an important role in nuclear structure physics, 
both continuous and discrete symmetries. Examples of continuous symmetries are isospin 
symmetry \cite{Heisenberg}, Wigner’s combined spin-isospin symmetry \cite{Wigner}, the 
(generalized) seniority scheme \cite{Racah,Talmi71}, the Elliott model \cite{Elliott} 
and the interacting boson model \cite{ibm}. Discrete symmetries have been used in the 
context of collective models to characterize the intrinsic shape of the nucleus, such 
as axial symmetry for quadrupole deformations \cite{BMVol2}, octupole \cite{octupole}, 
tetrahedral \cite{Dudek1,Dudek2} and octahedral \cite{Dudek2,Piet} symmetries for 
deformations of higher multipoles. 

A different application is found in $\alpha$-particle clustering in light nuclei to 
describe the geometric configuration of the $\alpha$ particles. Early work on $\alpha$-cluster 
models goes back to the 1930's with studies by Wheeler \cite{wheeler}, and Hafstad and Teller 
\cite{Teller}, followed by later work by Brink \cite{Brink1,Brink2} and Robson \cite{Robson}. 
Recently, there has been a lot of renewed interest in the structure of $\alpha$-cluster nuclei, 
especially for the nucleus $^{12}$C \cite{FreerFynbo}. The measurement of new rotational 
excitations of the ground state \cite{Fre07,Kirsebom,Marin} and the Hoyle state \cite{Itoh,Freer,Gai,Fre11} 
has stimulated a large theoretical effort to understand the structure of $^{12}$C ranging from 
studies based on the semi-microscopic algebraic cluster model \cite{Cseh}, 
antisymmetrized molecular dynamics \cite{AMD}, fermionic molecular dynamics \cite{FMD}, 
BEC-like cluster model \cite{BEC}, (no-core) shell models \cite{Roth,Draayer}, 
{\it ab initio} calculations based on lattice effective field theory \cite{Epelbaum1,Epelbaum2}, 
and the algebraic cluster model \cite{Marin,ACM,O16}. 

In this contribution, I discuss some properties of the $\alpha$-cluster nuclei $^{12}$C 
and $^{16}$O in the algebraic cluster model, and present some first results 
for odd-cluster nuclei in the framework of the cluster shell model.

\section{ALGEBRAIC CLUSTER MODEL}

The Algebraic Cluster Model (ACM) is an interacting boson model (IBM) to describe the relative 
motion of $n$-body clusters in terms of a spectrum generating algebra (SGA) of $U(\nu+1)$ where 
$\nu=3(n-1)$ represents the number of relative spatial degrees of freedom. In the IBM, $\nu$ 
denotes the five quadrupole degrees of freedom, leading to a SGA of $U(6)$.  
For the two-body problem the ACM reduces to the $U(4)$ vibron model \cite{cpl}, for 
three-body clusters to the $U(7)$ model \cite{ACM,BIL} and for four-body clusters 
to the $U(10)$ model \cite{O16,RB}. The ACM has a very rich symmetry structure. In addition 
to continuous symmetries like the angular momentum, in case of $\alpha$-cluster nuclei the 
Hamiltonian has to be invariant under the permuation of the $n$ identical $\alpha$ particles. 
Since one does not consider the excitations of the $\alpha$ particles themselves,  
the allowed cluster states have to be symmetric under the permutation group $S_n$.  

The potential energy surface corresponding to the $S_n$ invariant ACM Hamiltonian gives 
rise to several possible equilibrium shapes. In addition to the harmonic oscillator (or 
$U(3n-3)$ limit) and the deformed oscillator (or $SO(3n-2)$ limit), there are other 
solutions which are of special interest for the applications to $\alpha$-cluster nuclei. 
These cases correspond to a geometrical configuration of $\alpha$ particles located at the 
vertices of an equilateral triangle for $^{12}$C and of a regular tetrahedron for $^{16}$O. 
Even though they do not correspond to dynamical symmetries of the ACM Hamiltonian, one can 
still obtain approximate solutions for the rotation-vibration spectrum  
\ba
E &=& \left\{ \begin{array}{lll} 
\omega_{1}(v_{1}+\frac{1}{2}) + \omega_{2}(v_{2}+1) + \kappa \, L(L+1) && \mbox{ for $n=3$} \\ \\
\omega_{1}(v_{1}+\frac{1}{2}) + \omega_{2}(v_{2}+1) + \omega_{3}(v_{3}+\frac{3}{2}) + \kappa \, L(L+1) && \mbox{ for $n=4$} 
\end{array} \right. 
\label{energy} 
\nonumber
\ea
The rotational structure of the ground-state band depends on the point group symmetry of the 
geometrical configuration of the $\alpha$ particles and is summarized in Table~\ref{ACMsummary}. 

\begin{table}[h]
\centering
\caption{Algebraic Cluster Model for three- and four-body clusters} 
\label{ACMsummary}
\vspace{5pt}
\begin{tabular}{ccc}
\hline
\noalign{\smallskip}
& $3\alpha$ & $4\alpha$ \\
\noalign{\smallskip}
\hline
\noalign{\smallskip}
ACM & $U(7)$ & $U(10)$ \\
Point group & ${\cal D}_{3h}$ & ${\cal T}_{d}$ \\
Geometry & Triangle & Tetrahedron \\
\noalign{\smallskip}
\hline
\noalign{\smallskip}
G.s. band & $0^+$ & $0^+$ \\
          & $2^+$ &       \\
          & $3^-$ & $3^-$ \\
          & $4^{\pm}$ & $4^+$ \\
          & $5^-$ &       \\
          & $6^{\pm+}$ & $6^{\pm}$ \\  
\noalign{\smallskip}
\hline
\end{tabular}
\end{table}

\begin{figure}[h]
\begin{minipage}{.5\linewidth}
\includegraphics[width=\linewidth]{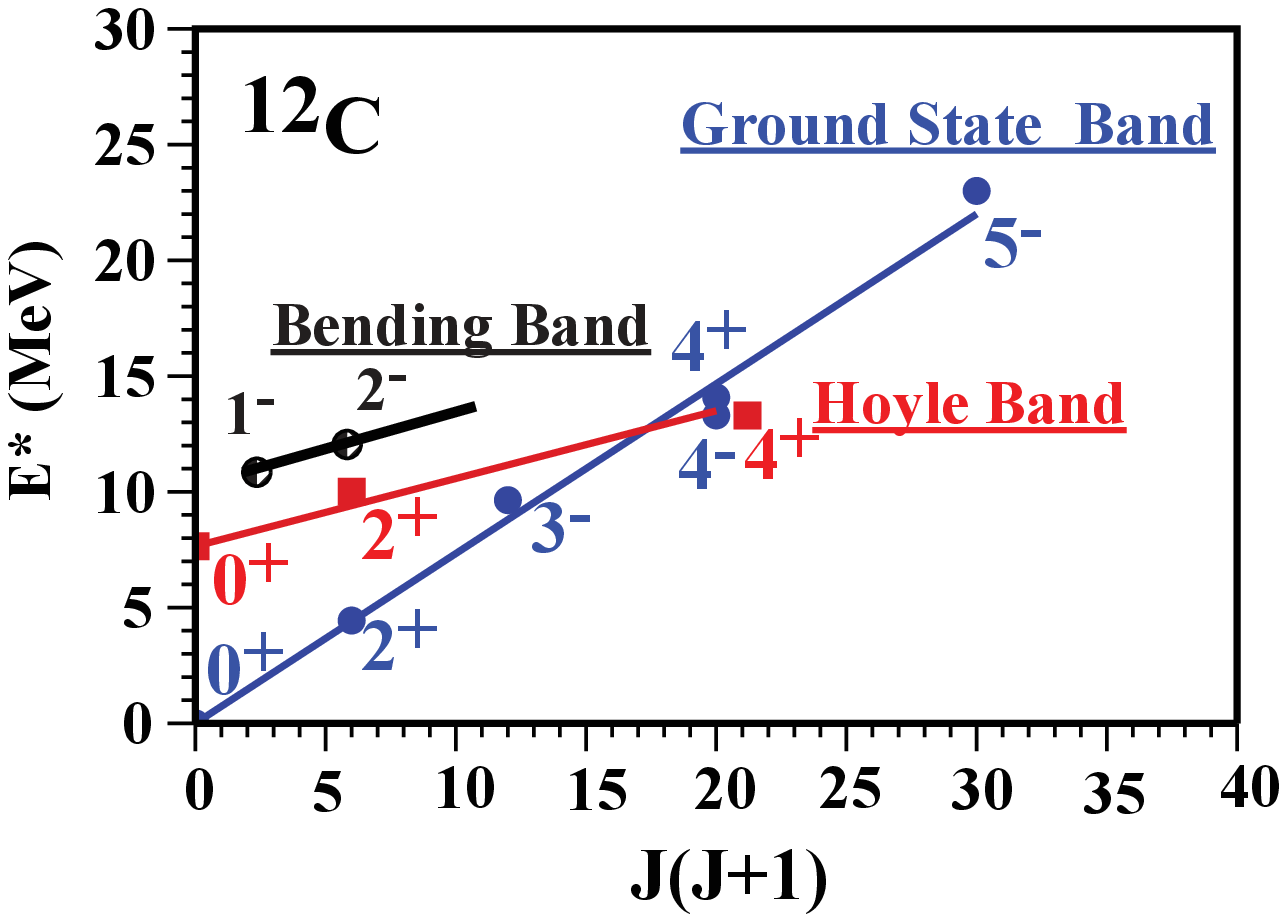}
\end{minipage}\hfill
\begin{minipage}{.5\linewidth}
\includegraphics[width=\linewidth]{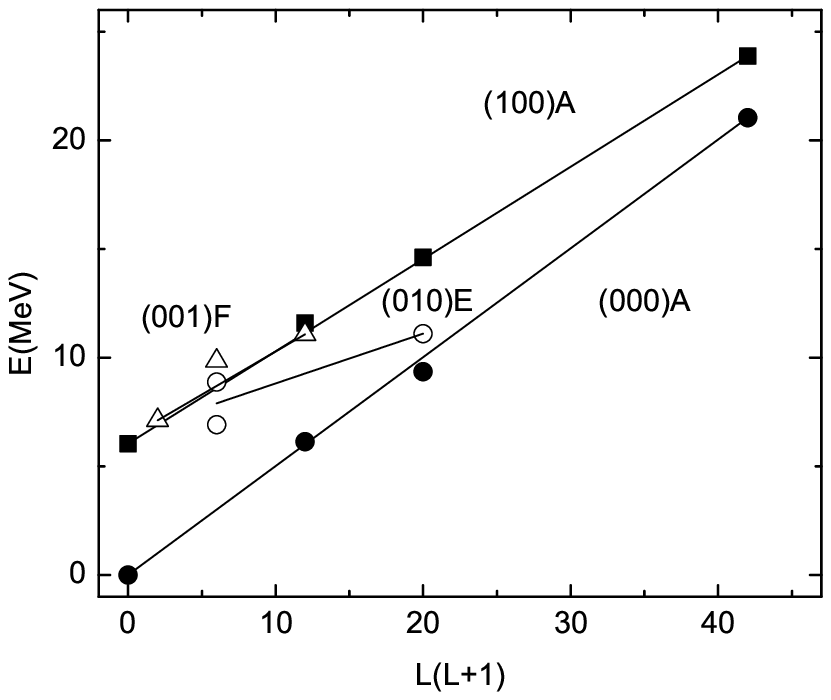}
\end{minipage}\hfill
\caption{(Color online) Rotational band structure of the ground-state band, the Hoyle band (or $A$ 
vibration) and the bending vibration (or $E$ vibration) in $^{12}$C (left) \cite{Marin}, 
and the ground-state band (closed circles), the $A$ vibration (closed squares), the $E$ vibration 
(open circles) and the $F$ vibration (open triangles) in $^{16}$O (right) \cite{O16}.}
\label{RotBands}
\end{figure}

The triangular configuration with three $\alpha$ particles has point group symmetry ${\cal D}_{3h}$ 
\cite{ACM}. Since ${\cal D}_{3h} \sim {\cal D}_{3} \times P$, the transformation properties under 
${\cal D}_{3h}$ are labeled by parity $P$ and the representations of ${\cal D}_{3}$ which is 
isomorphic to the permutation group $S_{3}$. The corresponding rotation-vibration spectrum is that 
of an oblate top: $v_{1}$ represents the vibrational quantum number for a symmetric stretching $A$ 
vibration, $v_2$ denotes a doubly degenerate $E$ vibration. The rotational band structure of $^{12}$C 
is shown in the left panel of Fig.~\ref{RotBands}. 

The tetrahedral group ${\cal T}_d$ is isomorphic to the permutation group $S_4$. In this case, there 
are three fundamental vibrations: $v_{1}$ represents the vibrational quantum number for a symmetric 
stretching $A$ vibration, $v_2$ denotes a doubly degenerate $E$ vibration, and $v_3$ a three-fold 
degenerate $F$ vibration. The right panel of Fig.~\ref{RotBands} shows the rotational band structure 
of $^{16}$O, where the bands are labeled by $(v_1,v_2,v_3)$. 

\section{ELECTROMAGNETIC TRANSITIONS}

For transitions along the ground state band the transition form factors are given in terms of a 
product of a spherical Bessel function and an exponential factor arising from a Gaussian distribution 
of the electric charges \cite{ACM} 
\ba
F(0^+ \rightarrow L^P;q) \;=\; c_L \, j_L(q \beta) \, \mbox{e}^{-q^{2}/4\alpha} ~.
\ea 
The transition form factors depend on the parameters $\alpha$ and $\beta$. 
The value of $\beta$ is determined from the first minimum in the elastic form factor 
giving $\beta=1.74$ (fm) \cite{ACM} for $^{12}$C and $\beta=2.07$ (fm) \cite{O16} for $^{16}$O. 
The value of $\alpha$ is determined by the size of the $\alpha$-particle 
to be $\alpha=0.56$ (fm)$^{-2}$ \cite{sick}. 

The transition probabilities $B(EL)$ along the ground state band can be extracted 
from the form factors in the long wavelength limit 
\ba
B(EL;0^+ \rightarrow L^P) \;=\; \frac{(Ze)^2}{4\pi} \, c_L^2 \, \beta^{2L} ~, 
\nonumber
\ea
with
\ba
c_L^2 = \left\{ \begin{array}{lcl} 
\frac{2L+1}{3} \left[ 1+2P_{L}(-\frac{1}{2}) \right] && \mbox{for $n=3$} \\ \\
\frac{2L+1}{4} \left[ 1+3P_{L}(-\frac{1}{3}) \right] && \mbox{for $n=4$} \end{array} \right.
\nonumber
\ea
This means that the transition probabilities $B(EL)$ for different multipolarities $L$ are related 
by the product of a numerical factor and a power of $\beta$,  
\ba
\frac{B(E3;3_{1}^{-} \rightarrow 0_{1}^{+})}{B(E2;2_{1}^{+} \rightarrow 0_{1}^{+})} 
&=& \frac{5}{2} \, \beta^2 ~,
\nonumber\\
\frac{B(E4;4_{1}^{+} \rightarrow 0_{1}^{+})}{B(E2;2_{1}^{+} \rightarrow 0_{1}^{+})} 
&=& \frac{9}{16} \, \beta^4 ~, 
\label{belc12}
\ea
for $^{12}$C, and 
\ba
\frac{B(E4;4_{1}^{+} \rightarrow 0_{1}^{+})}{B(E3;3_{1}^{-} \rightarrow 0_{1}^{+})} 
&=& \frac{7}{15} \, \beta^2 ~,
\nonumber\\
\frac{B(E6;6_{1}^{+} \rightarrow 0_{1}^{+})}{B(E3;3_{1}^{-} \rightarrow 0_{1}^{+})} 
&=& \frac{32}{45} \, \beta^6 ~, 
\label{belo16}
\ea
for $^{16}$O. 
The good agreement for the $B(EL)$ values for the ground band in Table~\ref{bem} shows 
that both in $^{12}$C and in $^{16}$O the positive and negative parity states indeed merge 
into a single rotational band. The large values of $B(EL;L_{1}^{P} \rightarrow 0_{1}^{+})$ 
indicate a collectivity which is not predicted for simple shell model states. 
The relation between octupole and quadrupole transitions in $^{12}$C, and that between 
octupole and hexadecupole transitions in $^{16}$O are beautifully explained by 
Eqs.~(\ref{belc12}) and (\ref{belo16}). 
The charge radii can be obtained from the slope of the elastic form factors 
in the origin, and are found to be in good agreement with the experimental value. 

The quadrupole moment for $^{12}$C can be calculated as 
\ba
Q_{2_1^+} \;=\; \frac{2}{7} \, Ze \, \beta^2 ~,
\ea
{\it i.e.} a positive value, as is to be expected for a oblate deformation, 
and is found to be in good agreement with a recent measurement \cite{kumar}.

\begin{table}
\centering
\caption{$B(EL)$ values and charge radii for $^{12}$C (top) and $^{16}$O (bottom). 
Experimental data are taken from \cite{ajz,reuter,strehl,kumar} and \cite{NDS}, respectively.}
\label{bem}
\vspace{5pt}
\begin{tabular}{cccl}
\hline
\noalign{\smallskip}
$^{12}$C & Th & Exp & \\
\noalign{\smallskip}
\hline
\noalign{\smallskip}
$B(E2;2_{1}^{+} \rightarrow 0_{1}^{+})$ &  8.4 & $7.6 \pm 0.4$ & $e^{2}\mbox{fm}^{4}$ \\
$B(E3;3_{1}^{-} \rightarrow 0_{1}^{+})$ & 73   & $103 \pm 17$  & $e^{2}\mbox{fm}^{6}$ \\
$B(E4;4_{1}^{+} \rightarrow 0_{1}^{+})$ & 44   &               & $e^{2}\mbox{fm}^{8}$ \\ 
$Q_{2_1^+}$ & $5.2$ & $5.3 \pm 4.4$ & $e\mbox{fm}^2$ \\ 
$\langle r^2 \rangle^{1/2}$ & 2.389 & $2.468 \pm 0.012$ & fm \\ 
\noalign{\smallskip}
\hline
\noalign{\smallskip}
$^{16}$O & Th & Exp & \\
\noalign{\smallskip}
\hline
\noalign{\smallskip}
$B(E3;3_1^- \rightarrow 0_1^+)$ &  215 & $205 \pm  10$ & $e^{2}\mbox{fm}^{6}$ \\
$B(E4;4_1^+ \rightarrow 0_1^+)$ &  425 & $378 \pm 133$ & $e^{2}\mbox{fm}^{8}$ \\
$B(E6;6_1^+ \rightarrow 0_1^+)$ & 9626 &               & $e^{2}\mbox{fm}^{12}$ \\ 
$\langle r^2 \rangle^{1/2}$ & 2.639 & $2.710 \pm 0.015$ & fm \\ 
\noalign{\smallskip}
\hline
\end{tabular}
\end{table}

\section{CLUSTER SHELL MODEL}

The next interesting question is how is the cluster configuration of $\alpha$-particles reflected in 
the neighboring odd-mass nuclei? What are the signatures of the underlying geometric symmetry in 
odd-cluster nuclei? Hereto the Cluster Shell Model (CSM) was developed \cite{DELLAROCCA2017158}. 
In this model nuclei with $Z=N=2n$ are treated as a cluster of $n$ $\alpha$-particles whose matter 
and charge density are given by a gaussian form
\ba
\rho\left(\vec{r}\right)=\left(\frac{\alpha}{\pi}\right)^{\frac{3}{2}}e^{-\alpha\left(r^{2}+\beta^{2}\right)}4\pi\sum_{\lambda\mu}i_{\lambda}\left(2\alpha\beta r\right)Y_{\lambda\mu}\left(\theta,\phi\right)\sum_{i=1}^{n}Y_{\lambda\mu}^{*}\left(\theta_{i},\phi_{i}\right).
\label{eq:distributionnucleus}
\ea
Here $\vec{r}_{i}=\left(\beta,\theta_{i},\phi_{i}\right)$ where $\beta$ denotes the relative distance of the $\alpha$ particles to the center of mass, and $\theta_{i}$ and $\phi_{i}$ are the angles. An important factor is the cluster factor, $\sum_{i=1}^{n}Y_{\lambda\mu}^{*}\left(\theta_{i},\phi_{i}\right)$, which contains the information about the geometrical configuration of the alpha particles.
\\
The potential is obtained by convoluting the density with a Volkov potential \cite{VOLKOV196533} to obtain
\begin{equation}
V\left(\vec{r}\right)=-V_{0}\sum_{\lambda\mu}4\pi e^{-\alpha\left(r^{2}+\beta^{2}\right)} i_{\lambda}\left(2\alpha\beta r\right)Y_{\lambda\mu}\left(\theta,\phi\right)\sum_{i=1}^{n}Y_{\lambda\mu}^{*}\left(\theta_{i},\phi_{i}\right)\label{eq:potencialcentral}.
\end{equation} 
The spin-orbit interaction is taken as
\begin{equation}
V_{so}\left(\vec{r}\right)=V_{0,so}\frac{1}{2}\left[\frac{1}{r}\frac{\partial V\left(\vec{r}\right)}{\partial r}\left(\vec{s}\cdot\vec{l}\right)+\left(\vec{s}\cdot\vec{l}\right)\frac{1}{r}\frac{\partial V\left(\vec{r}\right)}{\partial r}\right].
\end{equation}
Finally, for protons one has to include the Coulomb potential which is obtained by 
convoluting the carge density with the Green's function as 
\begin{eqnarray}
V_{C}(\vec{r}) &=& \frac{Ze^{2}}{n} \int \rho(\vec{r'}) \frac{1}{|\vec{r}-\vec{r'}|} d\vec{r'} 
\end{eqnarray}
We then obtain the single-particle energy levels and intrinsic states
as a function of $\beta$ for each configuration $\left(n=2,3,4\right)$
by solving the single-particle
Schr$\mathrm{\ddot{o}}$dinger equation
\begin{equation}
H=\frac{\vec{p}^{2}}{2m}+V\left(\vec{r}\right)+V_{so}\left(\vec{r}\right)+V_{C}\left(\vec{r}\right)\label{eq:hamiltonian free particle}.
\end{equation}
In the case of neutrons $V_{C}\left(\vec{r}\right)=0$.

\begin{figure}
\centering
\begin{minipage}{.5\linewidth}
\includegraphics[scale=0.6]{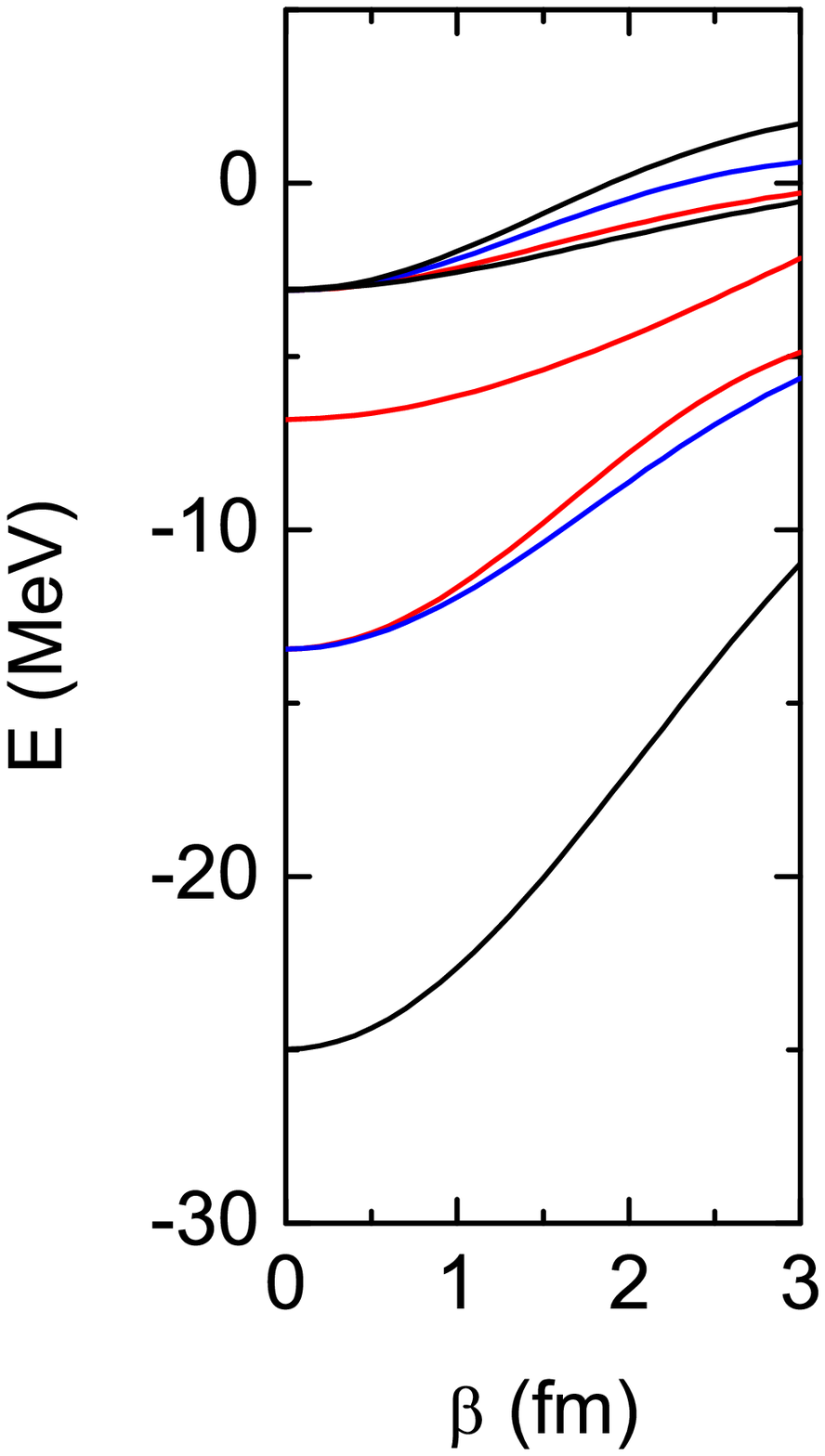}
\end{minipage}\hfill
\begin{minipage}{.5\linewidth}
\includegraphics[scale=0.6]{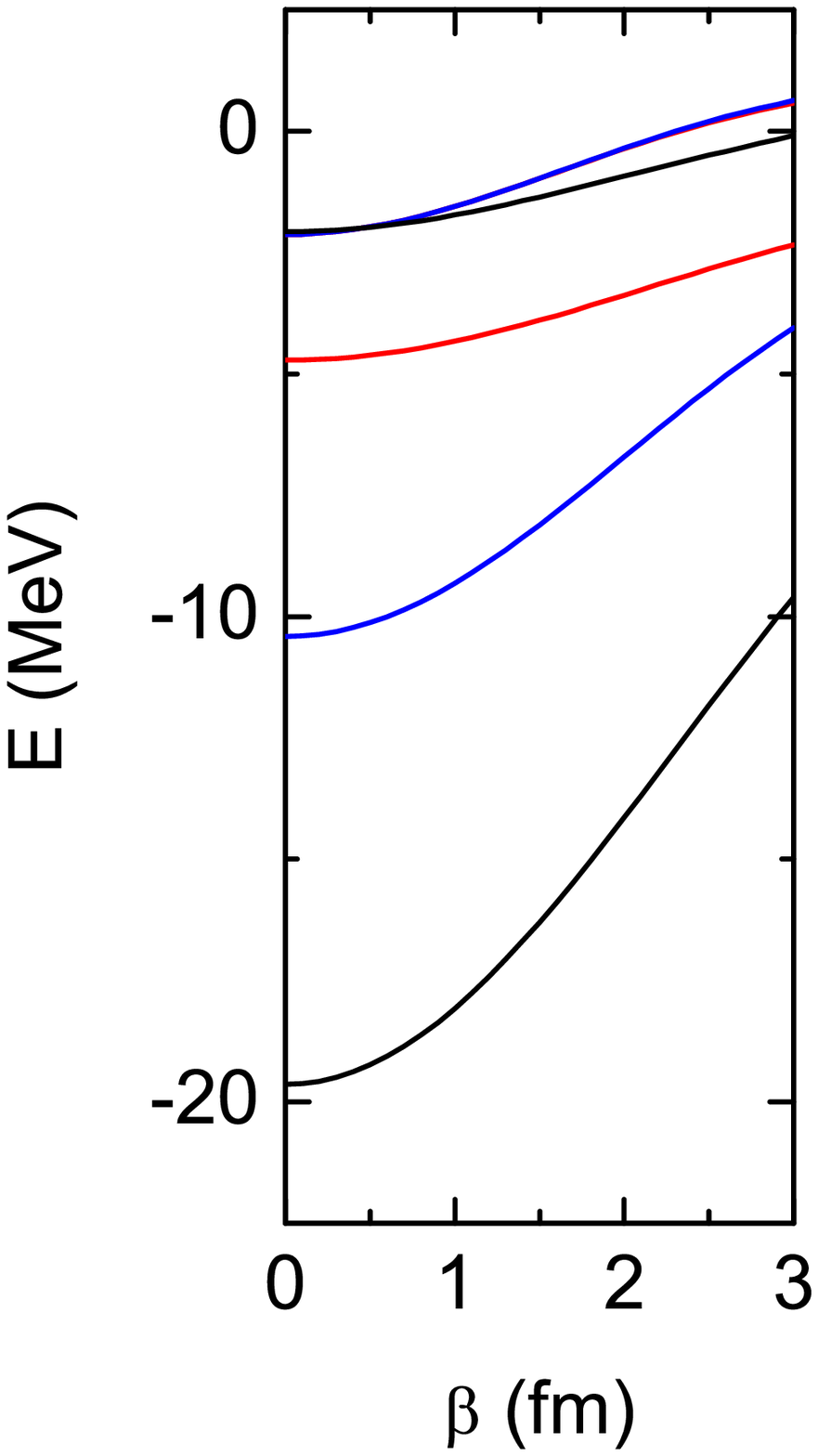}
\end{minipage}\hfill
\caption{Single-particle energies in a cluster potential 
with ${\cal D'}_{3h}$ triangular symmetry (left) and 
with ${\cal T'}_d$ tetrahedral symmetry (right). 
In the left panel, the single-particle levels are labeled 
by $E_{1/2}$ (black), $E_{5/2}$ (red) and $E_{3/2}$ (blue), 
and in the right panel by $E_{1/2}$ (black), $E_{5/2}$ (red) 
and $G_{3/2}$ (blue). For $\beta=0$ the ordering of the single-particle 
orbits is $1s_{1/2}$, $1p_{3/2}$, $1p_{1/2}$ and (almost degenerate) $1d_{5/2}$, $2s_{1/2}$.}
\label{splevels}
\end{figure}

Fig.~\ref{splevels} shows how the single-particle orbits are split as a function 
of the deformation parameter $\beta$, the relative distance of the $\alpha$ particles 
with respect to the center of mass. The single-particle levels can be classified 
according to the irreducible representations (irreps) of the corresponding double 
point group, $E_{1/2}$, $E_{5/2}$ and $E_{3/2}$ for the case of triangular symmetry  
(point group ${\cal D}'_{3h}$) and $E_{1/2}$, $E_{5/2}$ and $G_{3/2}$ for tetrahedral 
symmetry (point group ${\cal T}'_d$), in the notation of Herzberg \cite{herzberg}. 
The $E$ states are double degenerate and the $G$ states fourfold degenerate. 
In neither case, angular momentum and parity are good quantum numbers (with the exception 
for $\beta=0$). 

The construction of representations of the double groups $D_{3h}^{\prime}$ and 
$T_{d}^{\prime}$ was carried out for applications to molecular physics by Herzberg 
\cite{herzberg} and to crystal physics by Koster \textit{et al.} \cite{koster}. 
Here I discuss an application in nuclear physics \cite{C13}. 
Table~\ref{sporbits} shows the decomposition of single-particle orbits into irreps 
double point groups ${\cal D}'_{3h}$ and ${\cal T}'_{d}$. For example, both the $p_{3/2}$ 
and $d_{3/2}$ orbits transform as $G_{3/2}$ under ${\cal T}'_{d}$, whereas under  
${\cal D}'_{3h}$ the $p_{3/2}$ orbit splits into $E_{5/2}$ and $E_{3/2}$, and the 
$d_{3/2}$ into $E_{1/2}$ and $E_{3/2}$. 

\begin{table}
\centering
\caption{Splitting of single-particle orbits into representations of the 
double point groups ${\cal D}'_{3h}$ (left) and ${\cal T}'_{d}$ (right).}
\label{sporbits}
\vspace{15pt}
\begin{tabular}{l|ccc|ccc}
\hline
\noalign{\smallskip}
& & ${\cal D}'_{3h}$ & & & ${\cal T}'_{d}$ & \\
& $E_{1/2}$ & $E_{5/2}$ & $E_{3/2}$ & $E_{1/2}$ & $E_{5/2}$ & $G_{3/2}$ \\
\noalign{\smallskip}
\hline
\noalign{\smallskip}
$s_{1/2}$ & 1 & 0 & 0 & 1 & 0 & 0 \\
$p_{1/2}$ & 0 & 1 & 0 & 0 & 1 & 0 \\
$p_{3/2}$ & 0 & 1 & 1 & 0 & 0 & 1 \\
$d_{3/2}$ & 1 & 0 & 1 & 0 & 0 & 1 \\
$d_{5/2}$ & 1 & 1 & 1 & 0 & 1 & 1 \\
$f_{5/2}$ & 1 & 1 & 1 & 1 & 0 & 1 \\
$f_{7/2}$ & 2 & 1 & 1 & 1 & 1 & 1 \\
\noalign{\smallskip}
\hline
\end{tabular}
\end{table}

Fig.~\ref{rotbands} shows the rotational bands for the case of a 
single-particle coupled to a triangular configuration of $\alpha$ particles. 
The left panel shows the result for the ground-state band of even-cluster nuclei, 
as has been observed in $^{12}$C. The ground-state band consists of a series of $K$-bands 
with $K=3k$ ($k=0,1,2,\ldots,$) and angular momenta $L=0,2,4,\ldots,$ for $K=0$ 
and $L=K,K+1,K+2,\ldots,$ for $K \neq 0$. The parity is given by $P=(-1)^K=(-1)^k$. 

The results for the case of the coupling of a single-particle level with $E_{5/2}$, 
$E_{1/2}$ and $E_{3/2}$ symmetry to the ground-state of a triangular configuration 
with $A'_1$ symmetry is presented in the 2nd, 3rd and 4th panels of Fig.~\ref{rotbands}. 
The symmetry character under ${\cal D'}_{3h}$ is given by the products 
$A'_1 \otimes E_{5/2}=E_{5/2}$, $A'_1 \otimes E_{1/2}=E_{1/2}$, and 
$A'_1 \otimes E_{3/2}=E_{3/2}$, respectively. 
Just as for even-cluster nuclei, each representation for odd-cluster nuclei consists 
of a series of $K$-bands given by \cite{koster,C13}
\ba
E_{1/2} &:& K^P= 1/2^+, 5/2^-, 7/2^-, \ldots, 
\nonumber\\
E_{5/2} &:& K^P= 1/2^-, 5/2^+, 7/2^+, \ldots, 
\nonumber\\
E_{3/2} &:& K^P= 3/2^{\pm}, 9/2^{\pm}, \ldots,
\ea
with angular momenta $J=K,K+1,K+2,\ldots$.

An appliciation of this scheme to the rotational bands in $^{13}$C is in progress \cite{C13}. 

\

\

\begin{figure}[h]
\setlength{\unitlength}{0.5pt}
\begin{picture}(790,430)(0,-30)
\thicklines
\put (  0,-30) {\line(1,0){790}}
\put (  0,400) {\line(1,0){790}}
\put (  0,-30) {\line(0,1){430}}
\put (160,-30) {\line(0,1){430}}
\put (390,-30) {\line(0,1){430}}
\put (620,-30) {\line(0,1){430}}
\put (790,-30) {\line(0,1){430}}

\put ( 40,350) {$\bf D_{3h}: A'_1$}
\put (230,350) {$\bf D'_{3h}: E_{5/2}$}
\put (460,350) {$\bf D'_{3h}: E_{1/2}$}
\put (660,350) {$\bf D'_{3h}: E_{3/2}$}

\put ( 30, 60) {\line(1,0){20}}
\put ( 30,120) {\line(1,0){20}}
\put ( 30,260) {\line(1,0){20}}
\put ( 90,180) {\line(1,0){20}}
\put ( 90,260) {\line(1,0){20}}
\put ( 40, 10) {$\bf 0^+$}
\put (100, 10) {$\bf 3^-$}
\put ( 55, 55) {$\bf 0^+$}
\put ( 55,115) {$\bf 2^+$}
\put ( 55,255) {$\bf 4^+$}
\put (115,175) {$\bf 3^-$}
\put (115,255) {$\bf 4^-$}
\put (190, 60) {\line(1,0){20}}
\put (190, 90) {\line(1,0){20}}
\put (190,140) {\line(1,0){20}}
\put (190,210) {\line(1,0){20}}
\put (190,300) {\line(1,0){20}}
\put (250,140) {\line(1,0){20}}
\put (250,210) {\line(1,0){20}}
\put (250,300) {\line(1,0){20}}
\put (310,210) {\line(1,0){20}}
\put (310,300) {\line(1,0){20}}
\put (200, 10) {$\bf \frac{1}{2}^-$}
\put (215, 55) {$\bf \frac{1}{2}^-$}
\put (215, 85) {$\bf \frac{3}{2}^-$}
\put (215,135) {$\bf \frac{5}{2}^-$}
\put (215,205) {$\bf \frac{7}{2}^-$}
\put (215,295) {$\bf \frac{9}{2}^-$}
\put (260, 10) {$\bf \frac{5}{2}^+$}
\put (275,135) {$\bf \frac{5}{2}^+$}
\put (275,205) {$\bf \frac{7}{2}^+$}
\put (275,295) {$\bf \frac{9}{2}^+$}

\put (320, 10) {$\bf \frac{7}{2}^+$}
\put (335,205) {$\bf \frac{7}{2}^+$}
\put (335,295) {$\bf \frac{9}{2}^+$}
\put (420, 60) {\line(1,0){20}}
\put (420, 90) {\line(1,0){20}}
\put (420,140) {\line(1,0){20}}
\put (420,210) {\line(1,0){20}}
\put (420,300) {\line(1,0){20}}
\put (480,140) {\line(1,0){20}}
\put (480,210) {\line(1,0){20}}
\put (480,300) {\line(1,0){20}}
\put (540,210) {\line(1,0){20}}
\put (540,300) {\line(1,0){20}}
\put (430, 10) {$\bf \frac{1}{2}^+$}
\put (445, 55) {$\bf \frac{1}{2}^+$}
\put (445, 85) {$\bf \frac{3}{2}^+$}
\put (445,135) {$\bf \frac{5}{2}^+$}
\put (445,205) {$\bf \frac{7}{2}^+$}
\put (445,295) {$\bf \frac{9}{2}^+$}
\put (490, 10) {$\bf \frac{5}{2}^-$}
\put (505,135) {$\bf \frac{5}{2}^-$}
\put (505,205) {$\bf \frac{7}{2}^-$}
\put (505,295) {$\bf \frac{9}{2}^-$}
\put (550, 10) {$\bf \frac{7}{2}^-$}
\put (565,205) {$\bf \frac{7}{2}^-$}
\put (565,295) {$\bf \frac{9}{2}^-$}
\put (650, 90) {\line(1,0){20}}
\put (650,140) {\line(1,0){20}}
\put (650,210) {\line(1,0){20}}
\put (650,300) {\line(1,0){20}}
\put (710,300) {\line(1,0){20}}
\put (660, 10) {$\bf \frac{3}{2}^\pm$}
\put (675, 85) {$\bf \frac{3}{2}^\pm$}
\put (675,135) {$\bf \frac{5}{2}^\pm$}
\put (675,205) {$\bf \frac{7}{2}^\pm$}
\put (675,295) {$\bf \frac{9}{2}^\pm$}
\put (720, 10) {$\bf \frac{9}{2}^\pm$}
\put (735,295) {$\bf \frac{9}{2}^\pm$}
\end{picture}
\vspace{15pt}
\caption{Structure of rotational bands for a triangular configuration of $\alpha$ particles 
in even-cluster nuclei (first panel) and odd-cluster nuclei with $E_{5/2}$, $E_{1/2}$ and 
$E_{3/2}$ symmetry (second, third and fourth panel). Each rotational band is labeled by the 
quantum numbers $K^P$.}
\label{rotbands}
\end{figure}

\section{SUMMARY AND CONCLUSIONS}

In this contribution, I reviewed a study of the cluster states in $^{12}$C and $^{16}$O 
in the framework of the ACM in which they were interpreted as arising from the rotations 
and vibrations of a triangular and tetrahedral configuration of $\alpha$ particles, respectively. 
An analysis of both the rotation-vibration spectra and electromagnetic transition rates 
shows strong evidence for the occurrence of ${\cal D}_{3h}$ and ${\cal T}_{d}$ symmetry 
in in $^{12}$C and $^{16}$O, respectively. In both cases, the ground state band consist of 
positive and negative parity states which merge to form a single rotational band. This 
interpretation is validated by the observance of strong $B(EL)$ values. The rotational 
sequences can be considered as the fingerprints of the underlying geometric configuration 
(or point-group symmetry) of $\alpha$ particles. 

In the second part, I discussed the Cluster Shell Model which is very similar in spirit to 
the Nilsson model, in which the single-particle levels are split in the deformed field of 
the core nucleus. In the CSM, the levels are split in the deformed field generated by the 
cluster potential with triangular ${\cal D}_{3h}$ or tetrahedral ${\cal T}_{d}$ symmetry. 
As a first result, I presented the quantum numbers of the deformed single-particle levels 
and, for the case of $^{12}$C plus a nucleon, the angular momentum content of the rotational 
bands. The study of odd-cluster nuclei is currently in progress. The first results for 
$^{9}$Be and $^{9}$B have been published \cite{Be9}, and the systems with three- and four-$\alpha$ 
particles plus a nucleon (particle or hole) are underway. 

\section{ACKNOWLEDGMENTS}

This contribution was presented at a the Symposium 'Symmetries and Order', in honor of 
Franco Iachello on the occasion of his retirement from Yale University. The title of the 
Symposium is most fitting to describe Franco's legacy. His work is characterized by an 
uncanny ability to observe patterns, regularity, order and symmetries where most other 
people would simply see data and numbers. His work has inspired many generations of physicists 
all around the world and continues to do so. In appreciation of his support, guidance and 
inspiration during my entire career I wish Franco a happy and productive retirement! 

This work is the result of a collaboration between UNAM and Yale University, and has benefitted 
greatly from discussions with Franco, as well as with Valeria Della Rocca, Omar Alejandro 
D{\'{\i}}az Caballero and Adrian Horacio Santana Vald\'es. 
This work was supported in part by research grant IN109017 from PAPIIT-DGAPA.

\end{document}